\documentclass[sigconf]{acmart}

\usepackage{url}
\usepackage{pifont}
\newcommand{\cmark}{\ding{51}}
\usepackage{multirow} 

\usepackage{tikz}
\newcommand\copyrightnotice[1]{
	\begin{tikzpicture}[remember picture,overlay]
		\node[anchor=south,yshift=20pt] at (current page.south) {\fbox{\parbox{\dimexpr\textwidth-\fboxsep-\fboxrule\relax}{#1}}};
	\end{tikzpicture}
}

\AtBeginDocument{%
  \providecommand\BibTeX{{%
    \normalfont B\kern-0.5em{\scshape i\kern-0.25em b}\kern-0.8em\TeX}}}

\copyrightyear{2021}
\acmYear{2021}
\setcopyright{acmlicensed}\acmConference[ARES 2021]{The 16th International Conference on Availability, Reliability and Security}{August 17--20, 2021}{Vienna, Austria}
\acmBooktitle{The 16th International Conference on Availability, Reliability and Security (ARES 2021), August 17--20, 2021, Vienna, Austria}
\acmPrice{15.00}
\acmDOI{10.1145/3465481.3470111}
\acmISBN{978-1-4503-9051-4/21/08}

\begin{document}

\title[Finding Phish in a Haystack]{Finding Phish in a Haystack: A Pipeline for Phishing Classification on Certificate Transparency Logs}

\author{Arthur Drichel}
\authornote{Both authors contributed equally to this research.}
\email{drichel@itsec.rwth-aachen.de}
\affiliation{%
	\institution{RWTH Aachen University}
	\city{}
	\country{}
}
\author{Vincent Drury}
\authornotemark[1]
\email{drury@itsec.rwth-aachen.de}
\affiliation{
	\institution{RWTH Aachen University}
	\city{}
	\country{}
}
\author{Justus von Brandt}
\email{justus.von.brandt@rwth-aachen.de}
\affiliation{
	\institution{RWTH Aachen University}
	\city{}
	\country{}
}
\author{Ulrike Meyer}
\email{meyer@itsec.rwth-aachen.de}
\affiliation{
	\institution{RWTH Aachen University}
	\city{}
	\country{}
}

\renewcommand{\shortauthors}{Drichel and Drury, et al.}

\begin{abstract}
Current popular phishing prevention techniques mainly utilize reactive blocklists, which leave a ``window of opportunity'' for attackers during which victims are unprotected.
One possible approach to shorten this window aims to detect phishing attacks earlier, during website preparation, by monitoring Certificate Transparency (CT) logs.
Previous attempts to work with CT log data for phishing classification exist, however they lack evaluations on actual CT log data.
In this paper, we present a pipeline that facilitates such evaluations by addressing a number of problems when working with CT log data.
The pipeline includes dataset creation, training, and past or live classification of CT logs.
Its modular structure makes it possible to easily exchange classifiers or verification sources to support ground truth labeling efforts and classifier comparisons.
We test the pipeline on a number of new and existing classifiers, and find a general potential to improve classifiers for this scenario in the future.
We publish the source code of the pipeline and the used datasets along with this paper~\cite{ctl-pipeline}, thus making future research in this direction more accessible.
\end{abstract}

\ccsdesc[300]{Security and privacy~Intrusion detection systems}
\ccsdesc[300]{Computing methodologies~Machine learning}

\keywords{Phishing detection, certificate transparency, machine learning}


\maketitle
\section{Introduction}
\label{sec:introduction}
Despite a long history of research into phishing and its prevention, phishing attacks still pose a risk to internet users worldwide.
They impact the lives of general users, with more than 100,000 cases reported to the FBI in 2019~\cite{ic3_annual_2019}, and have been shown to be one of the most popular first steps in sophisticated attacks against larger organizations~\cite{verizon_dbir_2020}.

\copyrightnotice{\copyright\space Copyright held by the owner/author(s) 2021. This is the author's version of the work. It is posted here for your personal use. Not for redistribution. The definitive version was published in Proceedings of the 16th International Conference on Availability, Reliability and Security (ARES 2021), https://doi.org/10.1145/3465481.3470111}
Existing prevention techniques leave users vulnerable due to the gap between the start of an attack and its detection and subsequent inclusion of the correspondent URL in blocklists~\cite{oest_sunrise_2020}.
On the other hand, attackers are still evolving, changing their methods and tactics to keep ahead of defense technology.
Recently, a vast majority of phishing websites have started to offer valid HTTPS certificates to their users~\cite{apwg2020q3}.
Even though this trend leads to more legitimate-looking websites, it also opens an opportunity for researchers to detect phishing websites earlier in the process of a typical attack.
This is due to the fact, that when major browsers check a certificate's validity, they also require it to be present in so-called Certificate Transparency (CT) logs~\cite{ct_rfc_2013,scheitle_rise_2018}.
These public logs therefore offer a view of all certificates that are to be used by general users, which includes the certificates of phishing websites.
Certificates need to appear in at least two logs to be trusted by current major browsers~\cite{apple_ct, chrome_ct} and are often submitted directly by the issuer of a certificate when it is created~\cite{lets_ct}.
Logs are operated by several entities, including Google, Cloudflare and DigiCert, and can be monitored by anyone to make sure the log operators are working as expected.
Consequently, by monitoring these CT logs, it is possible to detect phishing websites while they are still being prepared by attackers.
This early detection can help shorten the gap between the start of a phishing attack and its inclusion in commonly used blocklists.

There are, however, several problems with this detection approach.
First, there is a large amount of certificates published in CT logs (see e.g.~\cite{li_certificate_2019}), making low numbers of false positives in potential classifiers a hard requirement.
Next, there is no ground truth available for all certificates in the CT logs, since there is no complete set of benign or malicious certificates that could be used to label the logs.
This makes it harder to train or test classifiers on the actual logs, as datasets are likely to be noisy and questions arise about the validity of performance measurements.

Consequently, prior work looking at the classification of certificates in CT logs exists, but lacks evaluations on real-world data (see Section~\ref{sec:related_work}).
In this paper, we present a modular pipeline that can be utilized to perform classifier analysis on CT logs.
The pipeline offers functionality from dataset creation, training of classifiers, up to real-world evaluation with variable verification sources to provide ground truth labeling.
The modular construction makes it easy to exchange and compare classifiers, or add additional verification sources.
The pipeline therefore offers a first step towards the evaluation of certificate classifiers on real CT log data, including the real-time detection of the certificates of phishing websites.

We use this pipeline in a preliminary comparison of several classifiers, and show that there is still a large potential for improvements, as none of the classifiers achieve satisfactory true positive rates while preserving manageable false positive rates.
These current shortcomings could be addressed in the future by removing noise from the training sets, as well as further improving the ground-truth labeling of CT log test data, tasks that are facilitated by the pipeline.
Finally, we publish the source code of the pipeline, as well as our datasets, in order to support more research in this area in the future~\cite{ctl-pipeline}.
It is our hope, that the public availability of the source code will make the creation and evaluation of classifiers on CT log data easier for the research community.

The remainder of this paper is structured as follows: We present related work in the next section, followed by the description of the proposed detection pipeline in Section~\ref{sec:detection_pipeline}.
Sections~\ref{sec:classifiers} and~\ref{sec:evaluation} include the description of the used classifiers and evaluation results, respectively.
Finally, we discuss our results in Section~\ref{sec:discussion} before concluding the paper.

\section{Related Work}
\label{sec:related_work}
The automatic detection of phishing websites has a long history, and consequently a large amount of work is available (see e.g.~\cite{das_sok_2020} for an overview).

In general, phishing attacks can be detected during the distribution of the ``bait'', e.g. by detecting phishing emails, or by inspecting the phishing websites themselves.
Different approaches work on different levels and can be combined, e.g. email detection prevents a phishing email from reaching the users' inbox, while website detection prevents them from entering their information once they already clicked on a link.
The detection of phishing certificates in CT logs aims at detecting phishing sites even before the ``bait'' is sent to the user, offering an additional layer of defense.
As certificates typically contain multiple domain names, and domain names also form a large part of a URL, the task of phishing URL detection and certificate detection somewhat overlap.
In the following we therefore first discuss existing approaches for phishing URL detection and then discuss approaches that make use of certificates.

The classification of URLs is a popular approach to automated phishing website detection.
For example, Zhao et al.~\cite{zhao_cost_2013} present an approach to train a classifier on only small imbalanced training sets, still achieving high classification accuracy.
In a second example, Whittaker et al.~\cite{colin_large_2010} present a classifier that includes features from URLs and website content, with the goal of using it to automatically update blocklists.
In fact, many of the URL-based solutions supplement URL features with features from additional sources, like WHOIS~\cite{whois} information or website content.
The inclusion of additional features may improve the classification performance and coverage, but is not always possible for certificates in the CT logs, as these certificates do not include full URLs.

Some of these classifiers also make use of information from the certificates of the website.
For example, Mohammad et al.~\cite{mohammad_predicting_2014} include the usage of HTTPS, as well as information about the certificate's issuer in their feature set.
The corresponding dataset has also been made public and been used in several additional studies. 

Compared to the classification of URLs, it is likely that certificates offer less information.
Even though there are several additional fields in certificates, they are often very similar or identical for certificates issued by the same issuer~\cite{drury_certified_2019}.
As such, the domain names embedded in certificates are the most promising factor for our use-case.
However, the domains in certificates include much less information than a complete URL, as they do not include path information at all, and sometimes even do not include all subdomains when wildcard certificates are used.
Even more problematic is the case of phishing websites hosted on compromised infrastructure, where the certificate was not requested with malicious intent.

Still, several approaches that focus on certificates only for phishing detection exist.
In 2015, Dong et al.~\cite{dong_beyond_2015} proposed a number of certificate features of phishing websites, including the existence, length, and relationship of different fields in certificates.
They compared a number of classifiers, trained on certificates collected directly from known phishing and benign websites between late 2012 and 2015, and found that random forest (RF) classifiers achieved the highest precision.
To our knowledge, the first proof of concept for using CT logs as basis for phishing website classification is the Phishing Catcher~\cite{phishing_catcher}, available at a GitHub repository.
As for peer-reviewed research, Scheitle et al.~\cite{scheitle_rise_2018} noted in 2018 that the CT logs might offer a new perspective for phishing detection.
In a preliminary look at the logs utilizing regular expressions, they find a large number of certificates (more than 125,000) that likely impersonate a small number of popular services, but do not include an in-depth analysis.
Torroledo et al.~\cite{torroledo_hunting_2018} train a phishing classifier on certificates only, aiming to detect differences in the legitimacy of phishing and non-phishing certificates.
Using a highly imbalanced dataset, they achieve a precision of around $90\%$, which we were not able to reproduce or verify in our experiments.
Fasllija et al.~\cite{fasllija_phish_2019} train a classifier on domain names extracted from full URLs, arguing that it would be able to perform classification on CT logs as well.
However, they do not include such an evaluation.
Recently, Sakurai et al.~\cite{sakurai_discovering_2020} proposed a classifier that is specifically created for the task of CT log classification.
The classifier is based on regular expressions extracted from known phishing websites, and achieves promising results on certificates collected from Censys~\cite{censys}.
However, the static logic based on regular expressions is neither able to detect new phishing campaigns with unknown domain name patterns, nor is it suitable for detecting spear-phishing campaigns, which do not create large amounts of similar domain names in the first place.

In this paper, we propose a pipeline which makes it possible to evaluate new and existing classifiers on actual CT log data, including the possibility of classifying certificates as soon as they are added to the logs.
For the comparison with previous works, we choose the Phishing Catcher as well as the classifier by Sakurai et al., as they work directly on certificates, and we were able to obtain the source code.
In addition, we also present and evaluate a number of feature-based and deep learning classifiers as an alternative to the existing approaches.

\section{Detection Pipeline}
\label{sec:detection_pipeline}

In this section, we present our detection pipeline which allows for comparatively evaluating phishing certificate detection classifiers. 
Researchers are the main target group for the pipeline. They profit from an accelerated development process for new detection methods and from the possibility to assess and compare different classifiers in a unified setting.
However, the pipeline can also be used right away with already proposed detection methods to identify phishing websites even before they are activated. 
The pipeline is able to perform retrospective analysis in addition to live classification of certificates published in the CT logs. 
All complex tasks, beginning from data acquisition (collecting certificates from CT logs as well as by crawling from well-known phishing websites), over data pre-processing (filtering and sanitizing), data labeling (as benign or phishing), classifier training, the classification itself, evaluating the classifiers' performance, up to the preparation of the final results are covered by our approach. 

In the following, we first specify our design goals and thereafter provide an overview of our modular pipeline.
The source code is written in Python and publicly available.

\subsection{Design Goals}
\label{sec:design_goals}
During the development of our pipeline we defined several design goals (DG) which aim to assure the convenient usability as well as the benefit of our approach. 

\subsubsection*{{DG1) Handle Data Processing}}
In order to train and assess the performance of a classifier and to analyze the outcome of an evaluation, our pipeline has to cope with different types of data obtained from various data sources. These sources provide data in various formats and must therefore be normalized before aggregation. In addition, the aggregated data may contain duplicates as well as mislabeled samples that need to be filtered out before use. Our first design goal therefore ensures that there is a means of data processing which enables the appropriate generation of training data and the evaluation of trained classifiers. In addition, due to the targeted live detection of phishing certificates, it is mandatory to provide a means for result validation since ground truth information cannot easily be acquired.

\subsubsection*{{DG2) Setup for Comparative Evaluations}}
Comparison of new approaches to established classifiers is an important part of the research process. As such, our second design goal is to ensure easy reusability by other researchers, such that our pipeline can be used for the comparative analysis of phishing certificate detection classifiers. Thereby, we contribute to the research community by enabling the assessment of newly developed or improved classifiers in a unified setting.

\subsubsection*{{DG3) Speed \& Scalability}}
As our pipeline has to process a large amount of certificates published in the CT logs, our third design goal is to ensure the possibility for real-time classification. Further, our approach shall enable an efficient retrospective analysis of large amount of data.

\subsubsection*{{DG4) Modularity \& Extendability}}
Since research in this field is advancing rapidly, the fourth design goal is to ensure that already developed parts of the pipeline can easily be exchanged. For instance, this is particularly important when new detection methods are proposed. Instead of developing a new pipeline around a newly developed detection method, we require that our approach provides clear interfaces that allow for an easy exchange of developed pipeline parts (e.g. the detection method). Moreover, implemented modules should be easily extendable and novel modules should be easily integratable into the pipeline flow.

\subsection{Pipeline Overview}
\begin{figure}[!t]
	\centering
	\includegraphics[width=1.0\linewidth]{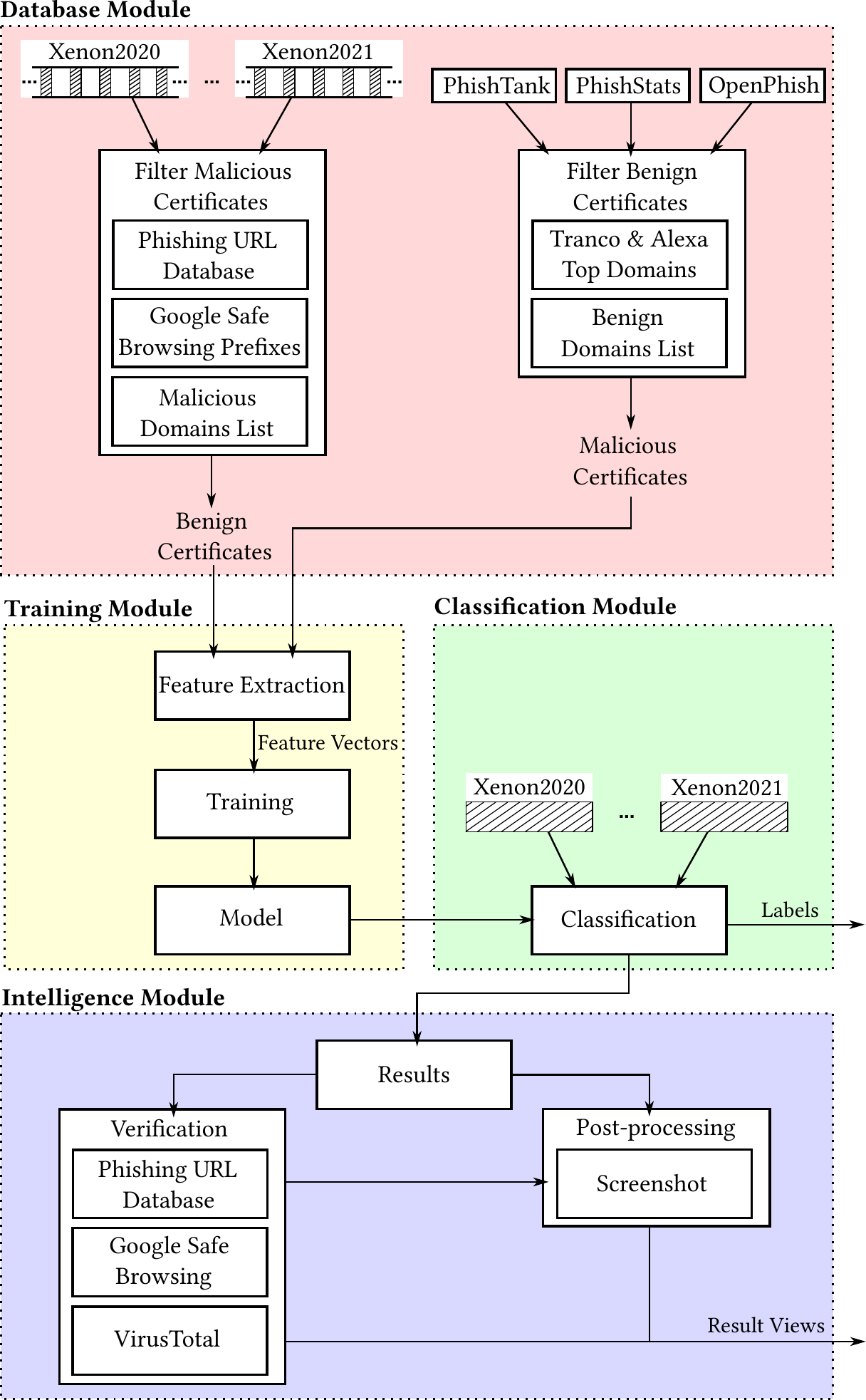}
	\caption{Abstract illustration of the architecture and pipeline operation.}
	\label{fig:architecture}
\end{figure}

Our detection pipeline consists of four distinct modules, each designed for a different task. We present an abstract illustration of the architecture including the pipeline operation in Fig.~\ref{fig:architecture} and describe each pipeline module in the following.

\subsubsection{Database Module}
\label{sec:database_module}
The database module is responsible for collecting, normalizing, sanitizing, labeling, and storing of data obtained from various data sources in a unified database. On the one hand, gathered data is used to train and evaluate different types of machine learning models. On the other hand, we utilize collected cyber threat intelligence to validate the results of live classifications retrospectively. Naturally, this information can also be used generally for retrospective analyses. We visualize this module in the upper part of Fig.~\ref{fig:architecture}. The outcome of the database module, in consideration to the pipeline flow, is a labeled dataset that can be used for classifier training.

Using the example of generating labeled training data that can be used for various types of machine learning models, we will now discuss the specific components of the database module in detail. We collect data from various sources to create a dataset, including benign and malicious labeled data, in order to enable supervised machine learning. 

We obtain benign certificates directly from the CT logs. Since the amount of certificates contained in these logs is very large, we divide the CT streams into chunks and define an API which allows for a convenient selection of benign certificates. In Fig.~\ref{fig:architecture}, we represent selected chunks by hatched boxes in the upper left part of the database module. We use chunks in order to obtain diverse and representative data that contains certificates published at different times of the day and different days of the week including working and non-working days.
The download process itself is highly parallelized. The number of downloader threads, the size of the chunks, the gap between the chunks and the time span to be taken into account can be freely selected. 

Since several major browsers require that a certificate has to be published in at least two different logs in order to be displayed without a warning, our download process might yield duplicate certificates. The database module thus includes a data sanitization process which removes duplicates. Additionally, we filter the benign certificates against a phishing URL database that includes known phishing URLs of the open source intelligence (OSINT) feeds of PhishTank~\cite{PhishTank}, PhishStats~\cite{PhishStats}, and OpenPhish~\cite{OpenPhish}.  Certificates usually contain more than one domain name,
 i.e., there are domains in the subject alternative name (\texttt{SAN}) field in addition to the common name (\texttt{CN}). We thus remove every certificate for which either the included CN or one of the SANs matches the domain name of a URL contained in our phishing URL database. 

Further, we  make use of Google Safe Browsing~\cite{GSB} which is a service based on two components for checking web resources on malicious content.  The first component contains a list of hash prefixes generated from malicious URLs.  The second component is an online service which can be queried if for a URL that is to be checked, the generated prefix is included within the first component. This two-component approach has the advantage of not disclosing the actual URL visited to anyone, while reducing the required number of online queries, as queries are only performed when a matching hash prefix is present.

We thus calculate for the CN and for every SAN the hash prefixes and remove all certificates for which we generate a collision with a hash included in the Google Safe Browsing prefixes. Note, since a hash collision is not a sufficient criterion for a URL being malicious, we might also remove non-malicious certificates of our benign labeled data with this procedure. However, due to large amount of available benign training samples we prefer to remove a few benign certificates over obtaining possibly contaminated data.  Moreover, this process saves us having to do any online queries.

We fetch new Google Safe Browsing prefixes as well as new phishing URLs from PhishTank and PhishStats every hour. Known phishing URLs from OpenPhish are fetched every twelve hours due to the restricted update frequency for the feed that is free of charge.\footnote{\url{https://openphish.com/phishing_feeds.html} online, accessed 2021-01-06}  In general, the update interval for the various data sources is adjustable and new data sources can easily be added as we have defined clear interfaces. For the already included cyber threat intelligence feeds we implement normalization methods in order to be able to store the data in a unified database, since every data source provides data in a different format.

Lastly, we permit the filtering of obtained benign certificates against a malicious domains list that is adjustable in order to enable the filtering of known malicious certificates that are not (yet) included in the observed cyber threat intelligence feeds. Currently, we do not make use of this filtering and thus our malicious domains list is empty. 

After these filtering steps we obtain benign labeled certificates which we use for the training of a supervised machine learning classifier.
 
We obtain malicious labeled data by downloading the certificates of phishing URLs that are contained in our phishing URL database using OpenSSL~\cite{OpenSSL}. Note, often redirecting services such as URL shorteners are used to hide the actual phishing URL.
We still decided against following redirects, opting to download the certificate of the URL shortening service instead, as this approach is less likely to be affected by website cloaking.
Website cloaking (see e.g.~\cite{oest2018inside}) is a method to evade detection by showing different versions of a website depending on, e.g. the geolocation of the request.
By requiring only the lower level TLS connection for the certificate download we minimize the effect of possible cloaking attempts.
In order to remove potential benign certificates from the set of malicious samples we filter against a benign domains list which contains common URL shorteners as well as common web hosting services. In total, we include 177 benign services to our filtering list after analyzing downloaded certificates and by using domain knowledge.

Additionally, we remove all duplicates and filter the CN as well as every SAN included in the malicious certificates using lists that include very popular domains. In detail, we filter against the top 1,000 Alexa~\cite{alexa} and top 1,000 Tranco~\cite{lepochat_tranco_2019} domains. The reason for this filtering is that it is unlikely that a phishing website is hosted on very popular domains.

With this procedure we are able to generate labeled training data. Note, the benign training data is only generated once and used for the training of machine learning classifiers. The gathering of cyber treat intelligence (e.g. for the phishing URL database) and the download of phishing certificates is, however, a continuous process. 
This is due to the fact that phishing websites often only have short lifetimes. Hence, we need to download the malicious certificates as soon as possible before the website is possibly removed.
In general, training data can be generated at will and used, for example, to improve an already trained classifier or to train a new one, as soon as enough new data has been collected.
Moreover, the database module can easily be extended by incorporating further filtering or by including additional cyber threat intelligence feeds.
The generated labeled training data is further processed by the training module which is described in the following.
 
\subsubsection{Training Module}
The training module is responsible for providing a trained model that is ready for live classification or retrospective analysis. It is illustrated in the left middle part of Fig.~\ref{fig:architecture}. Since this module is equipped with the generated labeled dataset from the database module, it is possible to train a classifier using supervised machine learning. However, this module also allows unsupervised machine learning models to be trained by simply ignoring the labels in the generated dataset. In addition, it is possible to define classification methods such as rule-based approaches that do not require any training at all. 

Further, it is possible to train feature-based as well as feature-less (i.e. deep learning based) approaches. Feature-based approaches, such as support vector machines (SVMs) or random forests (RFs), require the definition of features and the implementation of respective feature extraction methods. Contrarily, for feature-less approaches, such as recurrent (RNNs) or convolutional neural networks (CNNs), it is necessary to provide methods that encode all information that is relevant for classification such that it can be consumed by the neural network. In our case, such encoding methods could simply encode the characters of the CN and the SANs included in  a certificate via integer or one-hot encoding.

The result of the training module is a classifier which is ready to be used. We additionally serialize and store the classifier in order to enable result reproduction and the sharing of trained classifiers.

\subsubsection{Classification Module}
The classification module uses the trained model provided by the training module for classifying the certificates published in the CT logs. We display this module on the right middle part of Fig.~\ref{fig:architecture}. Here, it is possible to manually select a time span that should be taken into account for retrospective analysis or whether to perform live classification. Similar to the generation of the benign labeled training data in the database module, the respective CT logs that should be investigated can be selected freely. However, here we examine the full certificate streams instead of chunks. The output of the classification module are the predicted labels of the certificates published in the CT logs. These are output in real-time during live classification. Additionally, we pass the classification results to the intelligence module for further analysis.

\subsubsection{Intelligence Module}
The intelligence module receives the results passed by the classification module, tries to validate them, optionally performs post-processing, and processes the results into informative result views. We present the components of this module in the lower part of Fig.~\ref{fig:architecture}. 

One of the main tasks of this module is the verification of the results obtained from the classification module. Since we try to detect phishing websites even before they are activated, the classification results cannot be verified due to missing ground truth (i.e. when we detect a phishing certificate published in the CT logs in real-time it is unlikely that the corresponding phishing URL is already included in any OSINT feed). This is why the verification of live detection results is hardly possible. However, our verification process is particularly useful for research purposes and retrospective analyses. Moreover, the live classification results can be stored and later validated when new phishing URLs are added to the OSINT feeds. In detail, to verify classification results we make use of our offline phishing URL database, the Google Safe Browsing service, and VirusTotal\footnote{\url{https://www.virustotal.com/} online, accessed 2021-01-06}.

In addition, the intelligence module provides an interface which allows for easy post-processing of classification results.
Currently, we implemented a method which visits the domains contained in the CN and SAN fields for certificates that were labeled as positive by a classifier and makes a screenshot of the landing page.
These screenshots can then be examined manually or by an automatic approach to confirm the classification results.
Other post-processing approaches can also easily be added.
In particular, it might be possible to incorporate further automated filtering, which might reduce the number of false positives, but is too costly to perform for every domain in the CT logs.
Here, possible approaches are retrieving and analyzing the content of the websites, or obtaining intelligence from additional sources such as WHOIS~\cite{whois}.

Finally, the intelligence module calculates several metrics based on the results transferred by the classification module and presents them in aggregated results views. In addition, we perform a threshold analysis that can be used to determine the optimal operating point at any given false positive rate.

\section{Novel \& Existing Classifiers}
\label{sec:classifiers}

In this section, we introduce newly developed classifiers for certificate classification and present state-of-the-art classifiers proposed in related work.
We evaluate and compare the presented classifiers in Section~\ref{sec:evaluation} to test our proposed pipeline.
Apart from feature-based classifiers, we analyze deep learning classifiers, as well as two state-of-the-art approaches for phishing certificate detection.
For the features-based approaches we present our feature engineering and selection efforts which lay the basis for the developed machine learning classifiers.

\subsection{Classifying Domains}
\label{sec:classifiying_domains}
We tackle the phishing certificate detection task by dividing the classification of a single certificate into multiple domain name classification tasks and combining the results for a final decision.
Certificates usually include multiple domain names in the subject alternative name (\texttt{SAN}) field in addition to the common name (\texttt{CN}).
Since many of our envisioned features are to be extracted from domain names, we require a way to combine these separate domains into a single classification score for a given certificate.
We achieve this by performing individual predictions for each domain name included in a certificate and combining the results using a meta classifier.
We test several different meta classifiers: (a) maximum, (b) minimum, (c) average, and (d) median.
Each meta classifier uses the output of the domain classifiers and simply returns the maximum, minimum, average, or median value, respectively, as the classification result of the whole certificate.
Note, that the domain classifiers still utilize features from both domain and certificate, instead of only domain features.
We chose this simple approach over the alternative of classifying only information from the domains with the domain classifier, and using the certificate-specific features only in the meta classifier, as it does not require an additional training step.
We denote the combination of domain classifiers with a maximum, minimum, average, and median meta classifier by appending \textit{\mbox{-max}}, \textit{\mbox{-min}}, \textit{\mbox{-avg}}, and \textit{\mbox{-med}}, respectively, to the actual classification method.

\subsection{Feature Engineering \& Selection}
\label{sec:feature_engineering_and_selection}

To create a suitable feature set, we analyzed classifiers from related work \cite{bahnsen2017classifying, dong_beyond_2015, fasllija_phish_2019, schuppen_fanci_2018, torroledo_hunting_2018} and developed novel features by thoroughly analyzing benign and phishing certificates. We focus on context-less features, i.e. features that can be extracted from a single certificate, and omit features that are not relevant to our use-case (e.g. validity status of certificates, as CT logs only contain valid certificates). We do not obtain any intelligence from other sources (such as from WHOIS~\cite{whois}) in order to be independent of third party services and to ensure real-time classification capabilities.

Overall, the features we investigate can be split into three categories: (1) certificate-based features, (2) domain-based features and (3) keyword-based features. The first category contains features that are extracted from the certificate itself ignoring included domain names.
Instead, domain name specific features are included in the second category, except for the occurrence of specific keywords in the domain name, which are included in the third category.
We distinguish these feature sets as the list of keywords needs to be updated and is very specific, compared to the more general domain-based features.

The list of suspicious keywords is created from PhishTank URLs by analyzing parts of URLs split by dots after removing the public suffixes (top-level domains).
We sort the keyword candidates according to the number of their occurrences and reduce the list to 47 keywords by removing words that are too short or too general. We use each collected keyword for an individual feature which marks the presence of a keyword within a domain name. Additionally, we include two features which indicate the presence of any keyword and the total count of keywords in the domain.

In total, we gathered and engineered $126$ features, $22$ certificate-based features, $55$ domain-based features, and $49$ keyword-based features.
We present the full list of keywords that we have selected for the keyword-based features in Table~\ref{tab:keyword_features} in the Appendix.
Additionally, we list certificate-based and domain-based features in Table~\ref{tab:all_features} in the Appendix.
There, we present extracted feature values for a benign and a phishing certificate to make the features easily accessible.

After feature engineering, we define two different feature sets. The first feature set contains \textit{all} engineered features from all three categories and serves as baseline. The second feature set is created by performing feature selection on the features of the first two categories to ensure that only features that are actually relevant to the classification process are included. Additionally, reducing the number of features also reduces the required time for feature extraction and can improve a classifier by making it more robust to noise. We are limiting this feature set to a subset of features from the first two categories, thereby removing all keyword-based features. We argue, that this makes it more generally applicable, as it does not contain suspicious keywords, that cover specific targets and languages. For comparison, we evaluate classifiers using both feature sets, i.e. classifiers that make use of all 126 engineered features including keyword-based features and classifiers that only use a subset of domain- and certificate-features.

We perform feature selection by training an RF-based classifier and ranking its features by their importance according to the mean decrease in impurity (MDI)~\cite{gilles2013understanding}. We exclude features that are not important or exhibit high variances. This results in a total of $50$ features which belong to the first two feature categories. The \textit{selected} features are marked in Table~\ref{tab:all_features} in the Appendix. Note, feature selection is performed using a training dataset (see Section~\ref{sec:datasets}) that is completely disjoint to the actual test data used in our comparative evaluation.
In the following, we mark the utilized feature set of a classifier via an index (either \textit{all} or \textit{selected}). 

\subsection{Random Forest based Classifiers}
The first batch of classifiers we investigate on our proposed pipeline are random forest (RF) classifiers.
It has been shown that RF classifiers are well suited for the phishing website classification problem in the past (e.g. \cite{bahnsen2017classifying, dong_beyond_2015, sahingoz2019machine, subasi2017intelligent}).
For both feature sets (\textit{all} and \textit{selected}), we evaluate classifiers using all four meta classifier. This results in a total of eight classifiers that are to be evaluated. For all RF-based classifier we use the default hyperparameters (set by scikit-learn~\cite{scikit-learn}) but increase the number of estimators to $200$.

\subsection{Deep Learning based Classifiers}
To compare the feature-based approaches above to deep learning models, we create classifiers based on recurrent neural networks (RNNs).
RNNs have been shown to be suitable for URL classification tasks in several domains before (e.g. \cite{bahnsen2017classifying,torroledo_hunting_2018, woodbridge_predicting_2016}).

For neural network based classifiers, feature engineering and selection is not necessary. However, information that is relevant for classification has to be encoded and provided to the classifier. We thus utilize all engineered features as a sort of certificate encoding and provide this information to the model. In addition, we provide the raw domain names using characterwise integer encoding to the model. By choosing this approach, the neural network can (1) learn to extract relevant information from a domain name and (2) select the relevant information from all provided features by its own. 

\begin{figure}[!t]
	\centering
	\includegraphics[width=0.55\linewidth]{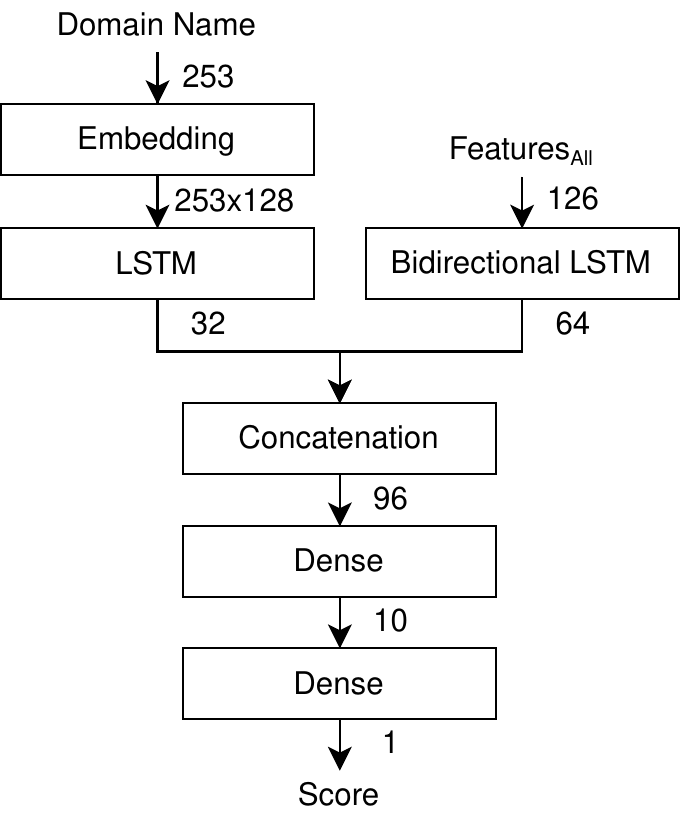}
	\caption{ Network architecture of the RNN-based classifier.}
	\label{fig:rnn_architecture}
\end{figure}

For the neural network classifier we choose an architecture in which the domain name and the features are consumed separately. We present the architecture of our RNN-based classifier including the input and output dimensions in Fig.~\ref{fig:rnn_architecture}. The input data is processed by distinct hidden layers and afterwards combined by concatenation for further processing. The final prediction is output by a fully connected layer.
In detail, the RNN-based approach uses one unidirectional long short-term memory (LSTM) layer for the domain name and one bidirectional LSTM layer to process the features. Before a domain name is fed into the model, we convert every included character to a unique integer and pad the result with zeros from the left side to the maximal domain length of 253 characters~\cite{mockapetris_domain_1987} as proposed in \cite{drichel_analyzing_2020}. This ensures that the model is able to process domain names at any length while using batch learning. The resulting encoded domain is processed by an embedding layer that adds additional information about the relationships between characters to the encoding. We choose an embedding dimension of 128 and thus project every character to a unique 128-dimensional vector. The embedded input is subsequently processed by an LSTM layer.

We also experimented with the RNN-based architecture proposed in \cite{torroledo_hunting_2018} but could neither reproduce the results stated in the original paper nor achieve better results than with our engineered architecture.

We optimized the neural network architecture iteratively using only data from the training set. As for the feature-based approach, this classifier also make use of meta classifiers, again using the maximum, minimum, average, and median variants.

\subsection{Classifying All Domains at Once}
In Section~\ref{sec:classifiying_domains}, we presented our approach for certificate classification by performing several domain classification tasks and combining the results using a meta classifier. In order to show that the certificate classification problem can also be solved using a single classification step and to demonstrate that different types of detection methods can be used in our pipeline, we here present a second approach that considers all domain names included in a certificate combined in one single classification step.

In this approach, we extract features for each individual domain included in a certificate and create a new feature vector that contains the average of the domain feature vectors for each feature. Additionally, depending on the used feature set, we also include certificate-based and keyword-based features. This makes it possible to combine variable numbers of domains into a single feature vector of constant length, an important requirement as the number of domains per certificate can vary greatly.
Further, this enables the use of the same architecture of classifiers for both approaches, as they use the same features.
We evaluate RF-based approaches using each defined feature set but do not implement a one-step certificate classifier approach for the deep learning model as we would need to average domain names.
We denote the approaches that classify the average of all feature vectors by appending \textit{\mbox{-cert}} to the actual classification method.

\subsection{State-of-the-Art Classifiers of Related Work}
In addition to the classifiers we newly developed, we evaluate two state-of-the-art approaches for phishing certificate detection proposed in related work.
We adapted both approaches slightly to comply with the interfaces defined in the training module.

\subsubsection{Sakurai}
The first classifier, by Sakurai et al.~\cite{sakurai_discovering_2020}, uses only domain names as input, and automatically creates regular expressions that match malicious domains following the same pattern.
This classifier does not return a classification score between $0$ and $1$, it only returns information about expressions that match the given domain.
We therefore modify the output to return the highest entropy reduction rate among all matching expressions (see~\cite{sakurai_discovering_2020}) as classification score.
Since the classifier is designed to find malicious domains, not certificates, we return the highest matching domain's score when combining the classification score for certificates, which corresponds to using the maximum meta classifier.

\subsubsection{Phishing Catcher}
Lastly, we evaluate \textit{Phishing Catcher}~\cite{phishing_catcher}, a rule-based \textit{certificate} classifier, where each rule potentially increases the classification score.
Rules include, for example, inclusion of suspicious keywords in any of the domains, or the usage of suspicious top-level domains.
The only non-domain specific feature indicates, whether the issuer of the certificate matches the popular free certificate authority ``Let's Encrypt''.
It is unique compared to the other classifiers in that it does not require any training, as it is based purely on heuristics based on domain knowledge.
We modify this classifier slightly to return a score between zero and one, instead of printing messages for suspicious domains.

\section{Evaluation}
\label{sec:evaluation}
In this section, we present our evaluation which is divided into two parts. In Section~\ref{sec:classifier_evaluation}, we evaluate and compare different approaches for distinguishing benign certificates from certificates created for phishing websites. Subsequently, in Section~\ref{sec:pipeline_evaluation}, we evaluate our proposed pipeline in consideration to the design goals defined in Section~\ref{sec:design_goals}.

\subsection{Classifier Evaluation}
\label{sec:classifier_evaluation}
First, we present the datasets utilized for the comparative evaluation of various phishing certificate detection classifiers. Thereafter, we present our evaluation method including the metrics that we observe. Lastly, we discuss the classifier evaluation results.

\subsubsection{Datasets}
\label{sec:datasets}
Below we describe the dataset we use for classifier training and provide an overview of the CT log data that we use for our comparative evaluation.

\paragraph*{Malicious labeled training data}
We use the database module (see Section~\ref{sec:database_module}) to generate malicious labeled training data. In detail, we obtain malicious certificates by downloading the PhishTank URL feed daily for the complete year of 2019. In addition, we collect further malicious certificates from the sources PhishTank and PhishStats by downloading their feeds hourly, and from OpenPhish by downloading its feed every twelve hours for the months January to May 2020. In total, we thereby collect 56,479 unique malicious certificates which we utilize for training classifiers.

\paragraph*{Benign labeled training data}
We obtain benign labeled training data also through our database module. Here, we download certificates from April 2020 of the Google Xenon~\cite{google_xenon} logs. These logs are some of the fastest advancing CT logs at time of writing. We download 70,889 unique benign certificates from which we randomly select 56,479 certificates (same number as malicious) for the training process.

We combine the collected benign and malicious certificates into a balanced training set that includes 112,958 certificates in total. In prior experiments, we tested different imbalanced data distributions in order to get closer to the actual distribution of the certificates in the CT logs. However, we could not measure a significant difference in performance between classifiers trained on imbalanced data. Thus, in the following we only use a balanced dataset to train the various classifiers. 

Moreover, we experimented with different sources for obtaining benign data. For instance, we downloaded certificates from popular websites according to Cisco Umbrella~\cite{cisco_umbrella}. However, classifiers trained on this data generally performed worse on a separate validation set containing actual CT log data. As such, we do not include them in our evaluation.

\paragraph*{CT log test data}
We chose the Google Xenon~\cite{google_xenon} logs for our comparative evaluation. As these logs are scoped, i.e. certificates that are included in Xenon2020 have an expiration date in 2020, we analyze all certificates published in the first week of May 2020 in Xenon2020, Xenon2021, Xenon2022, and Xenon2023. In total, these logs contain approximately 22.5 million unique certificates for the period under investigation. By selecting the first week of May as test data, we guarantee data chronology and disjoint training/testing data.

We released the utilized training dataset together with the source code for result reproducibility~\cite{ctl-pipeline}. Note, the used test data can easily be retrieved by our pipeline.

\subsubsection{Evaluation Overview}
We use the following software packages for our comparative evaluation: Python 3.7.3, scikit-learn 0.22, TensorFlow 2.3.0, Keras 2.4.3, CUDA 10.1, and cuDNN 7.6.5. The deep learning based approaches are executed on an NVIDIA Tesla V100 GPU. All other classification methods make use of Intel Xeon Platinum 8160
processors@2.1GHz. Note, our implementation is highly parallelized and all classification methods can be scaled with either more GPUs or with a greater number of CPUs (or CPU cores in general).

We train classifiers using the training dataset and evaluate them on the real-world CT log test data. The evaluation, and subsequent validation, was performed in December 2020, to ensure enough time has passed for malicious websites to be added to malicious domain feeds.
Note, that our evaluation approach is basically equivalent to performing the classifications in real time and verifying the results later, as the CT logs are append-only and therefore contain the same certificates in both settings.

In detail, we split a small portion from the training data for a validation set that is used either during the training in case of the deep learning based approaches, or after the training in case of RF-based approaches for model assessing purposes. We train the deep learning based models until there are no further improvement on the validation set for at least three consecutive epochs.

The Sakurai classifiers use only domain names as input and automatically create regular expressions that match malicious domains following the same pattern. 
We therefore use only one domain name from each certificate for training, namely the one that matches most closely the original URL from the malicious URL feed.
As the classifier distinguishes groups of domains by the number of dots present, we split our training set accordingly, and use up to 2,000 domains per group.
We train on all possible domain names for groups for which less than 2,000 samples are available.
For all other configuration options, we use the settings recommended in the original paper.

The Phishing Catcher classifier does not need any training at all since its classification relies on predefined rules.

We choose the false positive rate (FPR) and the true positive rate (TPR) as our evaluation metrics which are suitable measures especially for highly imbalanced data~\cite{davis2006relationship}. Since there is a far larger amount of benign certificates in the logs than phishing certificates, and the amount of certificates is large in general, we argue that a low FPR is the most important attribute of a suitable classifier. In addition, we observe the TPR which is a proxy for determining the amount of detected phishing certificates.

Note, the TPRs which we present are only estimates as we classify real-world data without any ground truth. We can thus only present a lower limit of the actual TPR because not every malicious certificate which is flagged positive by a classifier is verifiable via our verification process in the intelligence module. It is possible that there is no entry in our database for a correctly classified malicious certificate because the corresponding phishing website has not yet been reported to any of the observed OSINT feeds. This also implies that the FPRs presented may be slightly lower than stated.

\begin{figure}[!t]
	\centering
	\includegraphics[width=1.0\linewidth]{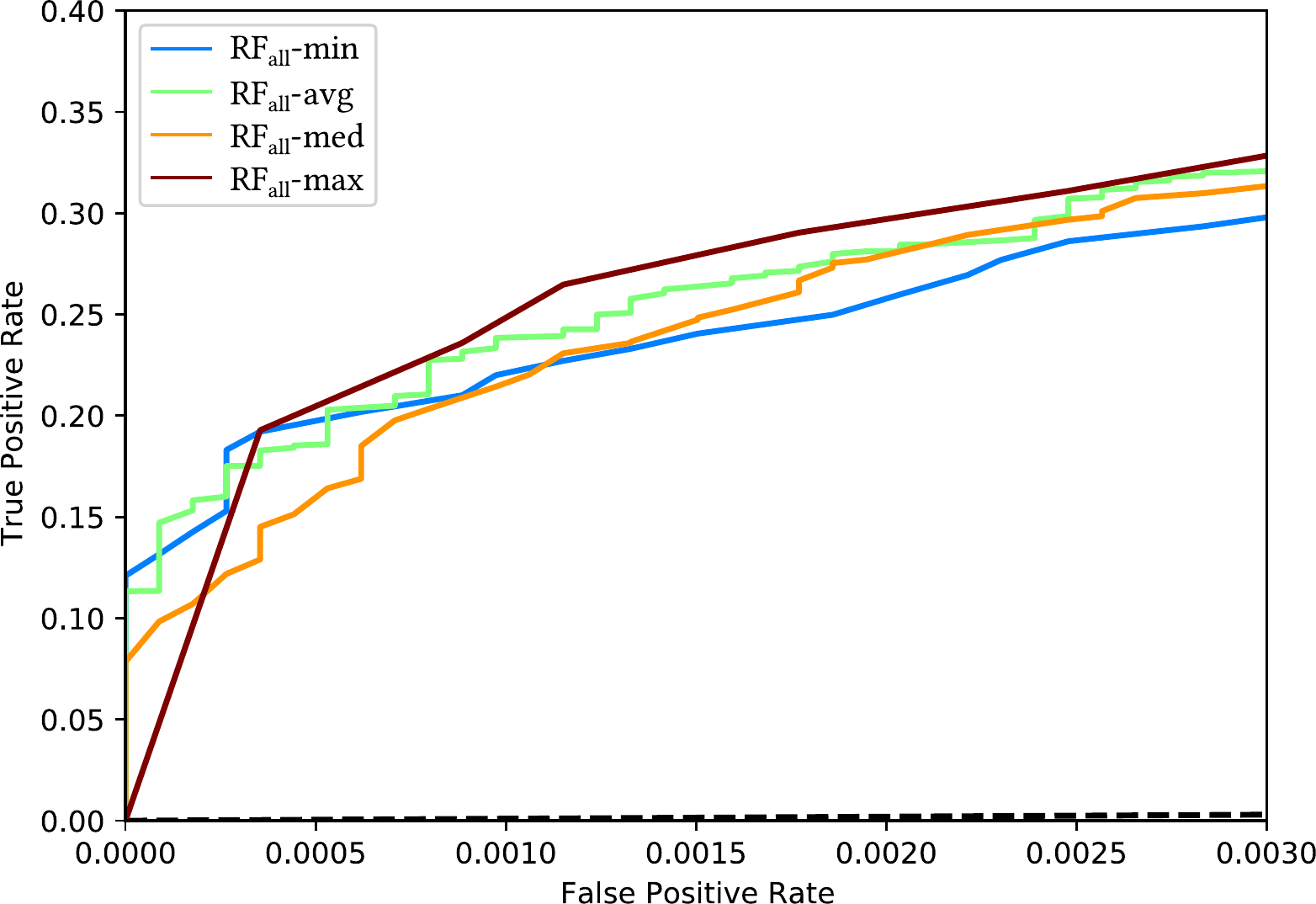}
	\caption{Model validation results: receiver operating characteristic curve of the RF\textsubscript{all} domain classifier using all four meta classifiers.}
	\label{fig:training_roc}
\end{figure}

As an example for model validation, in Fig.~\ref{fig:training_roc} we display the receiver operating characteristic curve obtained after training the domain RF\textsubscript{all} classifier (i.e. the RF-based domain classifier using all engineered features) for each of the four meta classifiers.

The receiver operating characteristic curve enables us to select an operating point which, in our further evaluation, is likely to produce very few false positive results.
In this diagram, we plot the x-axis finer as the FPR is the most important attribute of a classifier for our use-case.
The approaches which utilize the minimum and average meta classifier are promising and achieve a TPR of over 10\% at a very low FPR.
We display a baseline which corresponds to random guessing as a barely visible dashed line at the bottom of the diagram.
The baseline indicates that all four approaches work significantly better than random guessing.
While a TPR of around 10\% seems to be rather low, we argue that a classifier set at such an operating point can detect several phishing certificates without generating too many false positives.
Note, these are only model validation results. The actual results of the comparative evaluation on real-world CT log data are presented in the next subsection.

\subsubsection{Results}
\label{sec:classifier_results}

We present the results of the comparative evaluation in Table~\ref{tab:tpr-comparison}. Here, we display the total amount of detected phishing certificates and the estimated TPRs at fixed FPRs of $10^{-3}$ and $10^{-4}$ which correspond to a total of 22,510 and 2,251 false positives, respectively, for classifying the Google Xenon logs for a full week.

At the fixed FPR of $10^{-3}$, the rule-based classifier, Phishing Catcher, achieves by far the best results. 
At an FPR of $10^{-4}$, the classifier of Sakurai et al.\ detects the most phishing certificates followed by Phishing Catcher.
Our developed RF-based and RNN-based approaches achieve worse results. 
In general, the domain classifiers in combination with a meta classifier achieve better results compared to the one-step certificate classifiers which average the feature vectors.
The domain classifiers achieve better results using the minimum and average meta classifiers than by using the maximum or the medium meta classifiers.
The RF-based approaches which make use of all features (including keyword-based features) perform generally better than the classifiers that utilize the selected feature set.

We reckon the worse results obtained by the RF-based and RNN-based approaches compared to the classifier of Sakurai et al.\ and Phishing Catcher to be caused by noisy training data.
Since we obtained malicious labeled samples by downloading the certificates from URLs included in the various OSINT feeds, we also obtained benign certificates that were used on phishing websites hosted on compromised server infrastructure. Although, we filtered these certificates against our benign domains list, we reckon that a significant amount of benign certificates is falsely labeled as malicious as we only filtered against known redirecting and hosting services. According to estimates, $62\% - 73\%$ of phishing websites are actually hosted on compromised infrastructure \cite{le2019victim, le2019domain}. 

This is why the approaches by Sakurai et al. and Phishing Catcher perform slightly better than the machine learning classifiers. Phishing Catcher solely makes use of rules that were created by domain experts and therefore is not affected by compromised server certificates. The classifier proposed by Sakurai et al., on the other hand, classifies samples based on generated regular expressions that match malicious domains following the same pattern. Although, this classifier is trained on the same noisy data, it has a minor influence on the classification performance as we observe fewer patterns within the domain names of compromised server certificates.

The fact that the RF\textsubscript{all} classifiers achieve better results than the more general RF\textsubscript{selected} classifiers can be explained by the absence of keyword-based features within the selected feature set and by the fact that the selection was performed on the noisy training data.

With this evaluation, we were able to show that different types of machine learning classifiers can be used within the pipeline to detect phishing certificates. However, future research is needed as they are not working optimally.

\renewcommand{\arraystretch}{0.945}
\begin{table}[!t]
	\caption{Classifier Evaluation Results}
	\label{tab:tpr-comparison}
	\centering
		\begin{tabular}{lcccc}
			\toprule
			\multirow{2.5}{*}{\textbf{Classifier}} & \multicolumn{2}{c}{\textbf{FPR=0.001}} & \multicolumn{2}{c}{\textbf{FPR=0.0001}} \\
			\cmidrule(lr){2-3}
			\cmidrule(lr){4-5}
			& \textbf{\#TPs} & \textbf{TPR} & \textbf{\#TPs} & \textbf{TPR} \\
			\midrule
			RF\textsubscript{all}-min & 272 & 0.00418 & 55 & 0.00086\\
			RF\textsubscript{all}-avg & 243 & 0.00374 & 55 & 0.00086\\
			RF\textsubscript{all}-med & 232 & 0.00357 & 40 & 0.00062\\
			RF\textsubscript{all}-max & 244 & 0.00375 & -  & - \\
			\midrule
			RF\textsubscript{all}-cert & 145 & 0.00223 & -  & - \\
			\midrule
			RF\textsubscript{selected}-min & 216 & 0.00332 & 39 & 0.00060\\
			RF\textsubscript{selected}-avg & 208 & 0.00320 & 36 & 0.00055\\
			RF\textsubscript{selected}-med & 189 & 0.00291 & 29 & 0.00045\\
			RF\textsubscript{selected}-max & 148 & 0.00228 & -  & - \\
			\midrule
			RF\textsubscript{selected}-cert & 114 & 0.00176 & -  & - \\
			\midrule
			RNN-min & 352 & 0.00541 & 35 & 0.00054\\
			RNN-avg & 366 & 0.00563 & 36 & 0.00056\\
			RNN-med & 347 & 0.00533 & 34 & 0.00053\\
			RNN-max & 206 & 0.00317 & 20 & 0.00032\\
			\midrule
			Sakuarai et al. & 242 & 0.00373 & \textbf{117} & \textbf{0.00180}\\
			Phishing Catcher & \textbf{1102}  & \textbf{0.01692} & 110 & 0.00170\\
			\bottomrule
		\end{tabular} 
\end{table}
\renewcommand{\arraystretch}{1.0}

\subsection{Pipeline Evaluation}
\label{sec:pipeline_evaluation}
In the following, we evaluate the phishing certificate detection pipeline itself.
The goal of the pipeline creation was to build a system that can be used on real-world CT log data and accomplishes our design goals defined in Section~\ref{sec:design_goals}. Subsequently, we discuss whether we could achieve our envisioned design goals.

\subsubsection*{{DG1) Handle Data Processing}}
Our pipeline is able to collect, normalize, combine, filter, and label data from various data sources. This data is leveraged by our approach in order to train and evaluate classifiers. Moreover, we make use of the gathered intelligence to provide a means of result validation. We thus argue that our approach is able to handle complex data in different formats and from various data sources.

\subsubsection*{{DG2) Setup for Comparative Evaluations}}
In Section~\ref{sec:classifier_evaluation}, we comparatively evaluated several self-developed classifiers as well as classifiers that were proposed in related work. This shows that our approach enables the assessment of newly developed and improved classifiers in a unified setting. By providing this evaluation framework, we hope to bring the research community closer together as researchers can easily compare their developed classifiers with those proposed in related work.

\subsubsection*{{DG3) Speed \& Scalability}}
We performed the retrospective evaluation of a whole week of multiple CT logs (see Section~\ref{sec:classifier_evaluation}) in approximately one day. For instance, a single pipeline execution using RF classification with all four different meta classifiers was executed on 24 cores of an Intel Xeon Platinum 8160 processor@2.1 GHz, for which approximately 55 GB of RAM were required. We thus argue that our approach is able to process the large amount of certificates published in the CT logs as soon as they are added to the logs. Moreover, we made sure that our pipeline scales well with increasing computational power.

\subsubsection*{{DG4) Modularity \& Extendability}}
Extendability can hardly be evaluated by us. However, we designed our pipeline very modular and defined clear interfaces that allow for an easy integration of new modules. Moreover, we were able to show during our comparative evaluation that the training module can easily be exchanged.

\section{Discussion \& Future Work}
\label{sec:discussion}

The main focus of this work lies in the detection pipeline itself and not in developing the best phishing certificate detection classifiers.
In our comparative evaluation, we have shown that various types of classifiers, such as feature-based, deep learning based, and rule-based approaches, can be used inside our pipeline.
While our results show that it is possible to analyze several CT logs in real-time and to detect phishing certificates even before the corresponding phishing websites have been activated, the classifiers require future research.
By classifying the Google Xenon logs for a full week we could detect 117 malicious certificates while obtaining 2,251 false positives using the best evaluated classifier at a fixed FPR of $10^{-4}$.

While our approach enables the detection of certificates that are intentionally created for phishing websites, our approach is naturally not able to detect benign certificates used for phishing websites that are hosted on compromised server infrastructure. Future work could improve the filtering process during the creation of the malicious labeled training data in order to obtain less noisy data.
Unfortunately, the detection of certificates that are used on phishing websites hosted on compromised server infrastructure is likely not possible when the classification solely relies on information that is obtainable from a single certificate.
A further general restriction of using CT logs for phishing detection is, that potentially malicious subdomains can be removed from the certificate by using wildcard domain names.

As for the verification of certificates, our pipeline is currently not able to provide a complete ground truth for all certificates in CT logs.
The ground truth labeling depends on the usage of third-party repositories, such as PhishTank, Google Safe Browsing, or VirusTotal, that are neither guaranteed to be complete (i.e., contain all phishing websites), nor do they always provide information on newly created websites.
An alternative approach to the usage of third-party repositories would be to manually verify potential phishing websites.
Here, the pipeline is able to generate alerts for possible phishing websites, that can then be verified by an expert.
Contrary to the verification using external sources, we argue that manual verification is likely to be more effective when classifying the certificates of previously unseen domains. 
This is due to the fact that third-party repositories have a delay when adding new websites, while phishing websites usually have short lifetimes.
It is therefore unlikely that the third-party repositories include reliable information on all domain names that are added to the CT logs.
There are, however, also some problems with the manual verification approach.
First, many phishing websites use additional information in the URL path, that is not available from the certificate.
Guessing or otherwise acquiring this information might prove to be quite difficult, and might, together with using wildcard certificates, become an option for phishers to hinder the early detection that is offered by CT logs.
Further, the short lifetime of phishing websites implies the need to query potential phishing websites often, to be able to notice the transition from empty or placeholder page to the actual phishing website.
Even though we were able to detect and manually verify several previously unknown phishing websites in preliminary tests, the amount of empty or placeholder websites we encountered was far higher.
Finally, manual verification requires human interaction, and can therefore not be automated.

In this work, we have shown the feasibility of detecting phishing websites prior to their launch. However, an interesting study for future work is to analyze the profit in time that is obtainable by using our approach, i.e. to measure how long it takes until a phishing website appears in one of the observed OSINT feeds after the corresponding certificate was correctly detected by a classifier.

\section{Conclusion}
\label{sec:conclusion}
In this work, we presented a phishing certificate detection pipeline which allows for the classification of phishing certificates on CT logs in real-time, i.e., as soon as they are added to the logs.
Additionally, the pipeline enables retrospective analysis for research purposes and tackles the missing ground truth problem by providing a means of verification.
The pipeline developed is modular and extendable and provides a convenient framework for developing new certificate detection classifiers and comparing them with state-of-the-art approaches.
In addition, it enables the convenient generation of real-world data that can be used for training various types of machine learning classifiers.
By providing this framework, we hope to bring the research community closer together and to speed up future research.
We have evaluated the pipeline by performing a comparative evaluation of several self-developed classifiers and two state-of-the-art approaches proposed in related work showing the benefits of our approach.
We hope that this work brings the research community one step further towards closing  the phishers' window of opportunity.

\begin{acks}
	This project has received funding from the European Union's Horizon 2020 research and innovation programme under grant agreement No 833418.
	It was also supported by the research training group ``Human Centered Systems Security'' sponsored by the state of North-Rhine Westphalia.
	Simulations were performed with computing resources granted by RWTH Aachen University under project rwth0438.
	We are grateful to the authors of~\cite{sakurai_discovering_2020} for providing the source code of their classifier.
	We thank VirusTotal for providing us access to the Academic API.
\end{acks}

\bibliographystyle{ACM-Reference-Format}
\bibliography{bibliography}

\appendix
\section{Appendix}
\label{sec:appendix}

In Table~\ref{tab:keyword_features}, we present the full list of words we used for keyword-features.

\begin{table}[!h]
	\caption{Full list of words used for keyword-features.}
	\label{tab:keyword_features}
	\centering
	    \resizebox{\linewidth}{!}{
	\begin{tabular}{ccccc}
		\toprule
	secure
& login & mail & account & online  \\

	support & sites
& services & service & docs
\\
	update & signin & info
& security & help \\

	verify & recovery & mobile & secureserver
& storage \\

	center & verification & auth & promo & free \\
	paypal & runescape & google & apple & jppost \\
	sharepoint
& sagawa
& appleid
&	amazon
& icloud
\\
	windows & office
& facebook
& 1drv
& live
\\
	onedrive & ebay & allegro
&  itau
& bankofamerica
 \\
	cartetitolari & viabcp &&&\\
		\bottomrule
	\end{tabular}
		}
\end{table}

Table~\ref{tab:all_features} depicts extracted certificate and domain feature values for two example certificates, a benign and a phishing certificate. Additionally, we mark the features we selected during our feature selection process (Section~\ref{sec:feature_engineering_and_selection}).

\begin{table*}[!t]
	\caption{\textbf{Extracted certificate and domain features values for a benign ($\textbf{c\textsubscript{0}}$) and a phishing certificate ($\textbf{c\textsubscript{1}}$). A feature is defined as a function $\boldsymbol{\mathcal{F}}$ of a sample $\textbf{c}$. Domain features are extracted from common names (\textit{CNs}). \textit{CN}\textsubscript{$\textbf{c\textsubscript{0}}$} = \textit{anycast.ftl.netflix.com}, \mbox{\textit{CN}\textsubscript{$\textbf{c\textsubscript{1}}$} = \textit{paypal-secured.ga}}. Features selected during feature selection are marked. Categorical features: \textit{issuer}, \textit{key\_algorithm}}}
	\label{tab:all_features}
	\centering
	    \resizebox*{!}{\textheight}{
	\begin{tabular}{rlcllrr}
		\toprule
		\textbf{\#} & \textbf{Feature} & \textbf{Selected} & \textbf{Type} & \textbf{Output} & $\boldsymbol{\mathcal{F}}\textbf{(c\textsubscript{0})}$ & $\boldsymbol{\mathcal{F}}\textbf{(c\textsubscript{1})}$\\
		\midrule
		1           & is\_ov                                &        & certificate & binary   & 1       & 0       \\
		2           & is\_ev                                & \cmark & certificate & binary   & 0       & 0       \\
		3           & is\_dv                                &        & certificate & binary   & 0       & 1       \\
		4           & sub\_has\_c                           &        & certificate & binary   & 1       & 0       \\
		5           & sub\_has\_st                          &        & certificate & binary   & 1       & 0       \\
		6           & sub\_has\_l                           &        & certificate & binary   & 1       & 0       \\
		7           & sub\_only\_cn                         & \cmark & certificate & binary   & 0       & 1       \\
		8           & sub\_has\_cn                          &        & certificate & binary   & 1       & 1       \\
		9           & sub\_dn\_count                        &        & certificate & integer  & 6       & 1       \\
		10          & sub\_char\_count                      & \cmark & certificate & integer  & 64      & 17      \\
		11          & sub\_ext\_count                       &        & certificate & integer  & 10      & 9       \\
		12          & valid\_period                         & \cmark & certificate & integer  & 36      & 90      \\
		13          & policies\_count                       &        & certificate & integer  & 2       & 2       \\
		14          & is\_wildcard                          &        & certificate & binary   & 1       & 0       \\
		15          & has\_ocsp                             &        & certificate & binary   & 1       & 1       \\
		16          & has\_cdp                              & \cmark & certificate & binary   & 1       & 0       \\
		17          & san\_count                            & \cmark & certificate & integer  & 7       & 2       \\
		18          & average\_sd\_count                    & \cmark & certificate & rational & 4.14286 & 2.50000 \\
		19          & san\_tld\_count                       & \cmark & certificate & integer  & 2       & 1       \\
		20          & key\_algorithm                        &        & certificate & integer  & 2       & 1       \\
		21          & key\_size                             &        & certificate & integer  & 256     & 2048    \\
		22          & issuer                                & \cmark & certificate & integer  & 0       & 1      \\
		23          & sub\_cn\_entropy                      & \cmark & domain      & rational & 2.54753 & 2.47625 \\
		24          & sub\_cn\_is\_com                      & \cmark & domain      & binary   & 1       & 0       \\
		25          & name\_san\_entropy                    & \cmark & domain      & rational & 0.24027 & 0.08737 \\
		26          & has\_uppercase\_letters               &        & domain      & binary   & 0       & 0       \\
		27          & num\_dash                             & \cmark & domain      & integer  & 0       & 1       \\
		28          & num\_dash\_rd                         & \cmark & domain      & integer  & 0       & 1       \\
		29          & num\_tokens                           & \cmark & domain      & integer  & 4       & 3       \\
		30          & tld\_in\_token                        & \cmark & domain      & binary   & 1       & 0       \\
		31          & https\_in\_domain                     &        & domain      & binary   & 0       & 0       \\
		32          & longest\_token                        & \cmark & domain      & integer  & 7       & 7       \\
		33          & special\_char\_ratio                  & \cmark & domain      & rational & 0.13043 & 0.11765 \\
		34          & is\_ip                                &        & domain      & binary   & 0       & 0       \\
		35          & is\_idn\_domain                       & \cmark & domain      & binary   & 0       & 0       \\
		36          & san\_to\_alexa\_entropy               & \cmark & domain      & rational & 0.57761 & 0.74982 \\
		37          & vowel\_ratio                          & \cmark & domain      & rational & 0.23529 & 0.38462 \\
		38          & digit\_ratio                          & \cmark & domain      & rational & 0.00000 & 0.00000 \\
		39          & length                                & \cmark & domain      & integer  & 23      & 17      \\
		40          & contains\_wwwdot                      & \cmark & domain      & binary   & 0       & 0       \\
		41          & contains\_subdomain\_of\_only\_digits &        & domain      & binary   & 0       & 0       \\
		42          & subdomain\_lengths\_mean              & \cmark & domain      & rational & 5.66667 & 14.00000\\
		43          & parts                                 & \cmark & domain      & integer  & 3       & 1       \\
		44          & contains\_digits                      & \cmark & domain      & binary   & 0       & 0       \\
		45          & has\_valid\_tld                       &        & domain      & binary   & 1       & 1       \\
		46          & contains\_one\_char\_subdomains       & \cmark & domain      & binary   & 0       & 0       \\
		47          & prefix\_repetition                    &        & domain      & binary   & 0       & 0       \\
		48          & char\_diversity                       & \cmark & domain      & rational & 0.64706 & 0.78571 \\
		49          & contains\_tld\_as\_infix              & \cmark & domain      & binary   & 1       & 0       \\
		50          & alphabet\_size                        & \cmark & domain      & integer  & 11      & 11      \\
		51          & shannon\_entropy                      & \cmark & domain      & rational & 3.33718 & 3.37878 \\
		52          & hex\_part\_ratio                      & \cmark & domain      & rational & 0.00000 & 0.00000 \\
		53          & underscore\_ratio                     &        & domain      & rational & 0.00000 & 0.00000 \\
		54          & ratio\_of\_repeated\_chars            & \cmark & domain      & rational & 0.45455 & 0.27273 \\
		55          & consecutive\_consonant\_ratio         & \cmark & domain      & rational & 0.64706 & 0.14286 \\
		56          & consecutive\_digits\_ratio            & \cmark & domain      & rational & 0.00000 & 0.00000 \\
		57          & 1\_gram\_std                          & \cmark & domain      & rational & 0.65555 & 0.44536 \\
		58          & 1\_gram\_median                       & \cmark & domain      & integer  & 1       & 1       \\
		59          & 1\_gram\_mean                         & \cmark & domain      & rational & 1.54545 & 1.27273 \\
		60          & 1\_gram\_min                          &        & domain      & integer  & 1       & 1       \\
		61          & 1\_gram\_max                          & \cmark & domain      & integer  & 3       & 2       \\
		62          & 1\_gram\_bottom\_quartile             & \cmark & domain      & rational & 1.00000 & 1.00000 \\
		63          & 1\_gram\_top\_quartile                & \cmark & domain      & rational & 2.00000 & 1.50000 \\
		64          & 2\_gram\_std                          & \cmark & domain      & rational & 0.24944 & 0.27639 \\
		65          & 2\_gram\_median                       &        & domain      & integer  & 1       & 1       \\
		66          & 2\_gram\_mean                         & \cmark & domain      & rational & 1.06667 & 1.08333 \\
		67          & 2\_gram\_min                          &        & domain      & integer  & 1       & 1       \\
		68          & 2\_gram\_max                          & \cmark & domain      & integer  & 2       & 2       \\
		69          & 2\_gram\_bottom\_quartile             &        & domain      & rational & 1.00000 & 1.00000 \\
		70          & 2\_gram\_top\_quartile                & \cmark & domain      & rational & 1.00000 & 1.00000 \\
		71          & 3\_gram\_std                          & \cmark & domain      & rational & 0.00000 & 0.00000 \\
		72          & 3\_gram\_median                       &        & domain      & integer  & 1       & 1       \\
		73          & 3\_gram\_mean                         & \cmark & domain      & rational & 1.00000 & 1.00000 \\
		74          & 3\_gram\_min                          &        & domain      & integer  & 1       & 1       \\
		75          & 3\_gram\_max                          & \cmark & domain      & integer  & 1       & 1       \\
		76          & 3\_gram\_bottom\_quartile             &        & domain      & rational & 1.00000 & 1.00000 \\
		77          & 3\_gram\_top\_quartile                & \cmark & domain      & rational & 1.00000 & 1.00000 \\
		\bottomrule
	\end{tabular}
		}
\end{table*}

\end{document}